\documentclass[preprint]{aastex}
\usepackage{color}
\usepackage{epsfig}
\usepackage{graphicx}
\usepackage{amsmath}
\usepackage[english]{babel}
\usepackage[titletoc,title]{appendix}
\usepackage{txfonts}
\usepackage{times}
\usepackage{natbib}

\usepackage{hyperref}

\newcommand{\diffd}{\mathrm{d}}													
\newcommand{\equref}[1]{Eq.~\eqref{#1}}

\begin{document}

\title{
\vspace*{-2.0cm}
\hfill {\small \rm MPP-2016-321}\\[20mm]
\vspace*{-1.5cm}
Fundamental physics with the Hubble Frontier Fields: constraining Dark Matter models with the abundance of extremely faint and distant galaxies} 
\author{N.~Menci$^1$, A.~Merle$^2$, M.~Totzauer$^2$, A.~Schneider$^3$, A.~Grazian$^1$, M.~Castellano$^1$, N.G.~Sanchez$^4$ }
\affil{$^1$INAF -- Osservatorio Astronomico di Roma, via di Frascati 33, 00040 Monte Porzio Catone, Italy}
\affil{$^2$Max-Plank-Institut f\"ur Physik (Werner-Heisenberg-Institut), F\"ohringer Ring 6, 80805 M\"unchen, Germany}
\affil{$^3$Institute for Astronomy, Department of Physics, ETH Zurich, Wolfgang-Pauli-Strasse 27, 8093, Zurich, Switzerland}
\affil{$^4$LERMA, CNRS UMR 8112, 61, Observatoire de Paris PSL, Sorbonne Universit\'es, UPMC Univ.\ Paris 6, 61 Avenue de l'Observatoire, F-75014 Paris, France}
\begin{abstract}
We show that the measured abundance of ultra-faint lensed galaxies at $z\approx 6$ in the Hubble Frontier Fields (HFF) provides stringent constraints on the parameter space of  i)~Dark Matter models based on~keV sterile neutrinos; ii)~the ``fuzzy'' wavelike Dark Matter models, based on Bose-Einstein condensate of ultra-light particles. For the case of the sterile neutrinos, we consider two production mechanisms: resonant production through the mixing with active neutrinos and the decay of scalar particles. For the former model, we derive constraints for the combination of sterile neutrino mass $m_{\nu}$ and mixing parameter $\sin^2(2\theta)$ which provide the tightest lower bounds on the mixing angle (and hence on the lepton asymmetry) derived so far by methods independent of baryonic physics. For the latter we compute the allowed combinations of the scalar mass, its coupling to the Higgs field, and the Yukawa coupling of the scalar to the sterile neutrinos. We compare our results to independent, existing astrophysical bounds on sterile neutrinos in the same mass range. For the case of "fuzzy" Dark Matter, we show that the observed number density $\approx 1/{\rm Mpc}^3$ of high-redshift galaxies in the HFF sets a lower limit $m_\psi\geq 8\cdot 10^{-22}$~eV (at 3-$\sigma$ confidence level) on the particle mass, a result that strongly disfavors wavelike bosonic Dark Matter as a viable model for structure formation.  We discuss the impact on our results of uncertainties due to systematics in the selection of highly magnified, faint galaxies at high redshifts.
\end{abstract}
\section{\label{sec:Introduction}Introduction}
Understanding the nature of the Dark Matter (DM) component of the Universe constitutes a key issue in fundamental physics and in cosmology. During the last two decades, the formation and the growth of cosmic structures has progressively adopted the Cold Dark Matter (CDM) paradigm as a baseline~\citep{Peebles:1982ff,Blumenthal:1984bp}. This envisages DM particles to be characterized by thermal velocities small enough to produce negligible free streaming on the scales relevant to structure formation. Typically, this corresponds either to assuming DM particles to be massive ($m_X>0.1$~GeV) or to be constituted by condensates of light axions (with mass $\sim 10^{-5}-10^{-1}$~eV). Basic motivations for such a scenario are its simple properties (the corresponding power spectrum has a self-similar power law behavior in the whole range  of mass scales involved in galaxy formation) and the fact that particles with mass and cross sections characteristic of the weak scales (weakly interacting massive particles, WIMPs, with masses $m_X\sim 100$~GeV) produce approximately the correct abundance of DM when they freeze-out of equilibrium in the early Universe. On the other hand, extensive studies of structure formation showed that the CDM model provides an excellent baseline to explain the properties of large-scale structures and of galaxies on a huge range of mass scales ranging from $M\approx 10^{16}$ to $M\approx 10^9\,M_\odot$~\citep[see, e.g.,][for a review]{Diemand:2009bm}. 

However, as of now, both direct~\citep[see, e.g.,][]{Aprile:2012nq,Aprile:2015uzo,Akerib:2013tjd} and indirect~\citep[see, e.g.,][]{Adriani:2013uda,Ackermann:2015zua} CDM detection experiments have failed to provide a definite confirmation of such a scenario. Also, no evidence for CDM candidates with mass $10^2-10^4$~GeV has been found in experiments at LHC~\citep[e.g.,][]{ATLAS:2012ky}, while experiments aimed to detect axions as DM components have produced no evidence in the explored portion of the parameter space~\citep[see][]{Graham:2015ouw,Marsh:2015xka}. On the structure formation side, several critical issues are affecting the CDM scenario at the mass scales of dwarf galaxies ($M\approx 10^7-10^9$ $M_\odot$). These are all connected to the excess of power in the CDM power spectrum at such scales compared to a variety of observations, and include: the excess DM density of the inner regions of dwarf galaxies compared to the observed cored profiles~\citep[see][]{deVega:2013jfy}, the over-abundance of faint dwarfs around our Galaxy and in the Local Group~\citep[see, e.g.,][]{Lovell:2011rd} as well as in the field~\citep{Menci:2012kk,Maccio:2012qf,Papastergis:2014aba}, the excess of massive satellite DM halos with virial velocities $V_{\rm vir} \geq 20$ km/s relative to the number of observed bright dwarf galaxies~\citep{BoylanKolchin:2011de}, and -- most of all -- the over-prediction of the abundance of field dwarfs with $V_{\rm vir} \approx $ 40--60~km/s~\citep{Klypin:2014ira}.  

While a refined treatment of baryonic effects entering galaxy formation (in particular feedback from supernovae) has been proposed to solve or at least alleviate the above problems (see, e.g., Governato et al. 2012, 2015; Di Cintio et al. 2014), the combination of astrophysical issues with the lack of direct or indirect detection of candidate particles has stimulated the interest toward different DM scenarios, characterized by power spectra with suppressed amplitude at small mass scales ($M\lesssim 10^8-10^9$ $M_\odot$) with respect to the CDM case. In fact, several groups have started to investigate galaxy formation in a number of alternative models, such as self-interacting DM~\citep{Rocha:2012jg,Vogelsberger:2014pda}, decaying DM~\citep{Wang:2014ina}, late-forming DM~\citep{Agarwal:2014qca}, atomic DM~\citep{CyrRacine:2012fz}, and DM interacting with dark radiation~\citep{Buckley:2014hja,Chu:2014lja}.
 
In this framework, a specific focus has been given to  Warm Dark Matter (WDM) scenarios, which assume DM to be composed by particles with masses $m_X$ in the~keV range. Their larger thermal velocities (corresponding to larger free-streaming lengths) suppress structure formation at scales $M=10^7$--$10^9$ $M_\odot$, depending on the exact value of $m_X$ (since a thermalized species has no memory of the details of its production). While  WDM candidates may result from the freeze-out of particles initially in thermal equilibrium in the early Universe~\citep[like, e.g., gravitinos, see][for a review]{steffen2006}, a similar suppression at these scales can be obtained by a variety of models featuring particles in the~keV mass range with \emph{non-thermal} spectra, like sterile neutrinos. Note that, in the case of non-thermal spectra, the production mechanism is essential in determining the suppression of the power spectra with respect to CDM. The shape of the power spectra are -- although even that only qualitatively -- somewhat similar to thermal WDM cases only for specific regions of the parameter space of the assumed production model,  since a non-thermal spectrum cannot be associated with a temperature straightforwardly. Thus, to provide accurate limits to the mass of a non-thermal candidate, a detailed exploration of the parameter space of the selected models has to be performed. In this work, we tackle this task focusing on models of~keV sterile neutrinos, which received a particular interest in the literature in recent years~\citep{Adhikari:2016bei}. This is due to both solid fundamental physics motivations (right-handed neutrinos constitute a natural extension of the Standard Model to provide mass terms for active neutrinos, see~\cite{Merle:2013gea} and to the fact that such particles constitute the simplest candidates~\citep[see, e.g.,][]{Abazajian:2014gza} for a Dark Matter interpretation of the  potential X-ray line in stacked observations of galaxy clusters and in the Perseus cluster~\citep{Bulbul:2014sua,Boyarsky:2014jta}. In fact, in the presence of a tiny admixtures $\sin^2(2\theta)$ with the active neutrinos, the decay of sterile neutrinos can result into photon emission at energies close to $m_{\nu}/2$. The non-detection of such a line in galaxies, dwarf galaxies, or the Milky Way~\citep[see, e.g.,][for an extended discussion]{Sekiya:2015jsa,Jeltema:2015mee,Riemer-Sorensen:2014yda,Adhikari:2016bei}  yields effective \emph{upper} limits on the mass of sterile neutrinos for each value of the mixing angle in the range $10^{-13}\leq \sin^2(2\theta)\leq 10^{-9}$~\citep{Canetti:2012kh,Tamura:2014mta,Ng:2015gfa}, which rule out sterile neutrino models based on the non-resonant production mechanism by Dodelson-Widrow~\citep{Dodelson:1993je}.\footnote{Note that, contrary to the claims tracing back to the early references~\citep{Dodelson:1993je,Colombi:1995ze}, the spectrum resulting from non-resonant production is \emph{non-thermal}~\citep{Merle:2015vzu}, rather than being proportional to a Fermi-Dirac distribution multiplied by a suppression factor.} 

However, general \emph{lower} mass bounds (arising from phase space density of nearby dwarf galaxies) are rather loose, yielding a model-independent limit $m_{\nu}\geq 0.4$~keV~\citep{Boyarsky:2008ju}, and leave a major portion of the parameter space largely unconstrained, in scenarios in which sterile neutrinos are resonantly produced (RP~models) in the presence of a lepton asymmetry $L$ in the background medium~\citep[as proposed by][]{Shi:1998km}. They can, however, be significantly tightened when the information available from the production mechanism is taken into account~\citep{Schneider:2016uqi,Merle:2014xpa}, thereby strongly pushing such scenarios. Other settings, where sterile neutrinos are produced by decays of, e.g., scalar particles~\citep[see a few concrete examples in][] {Shaposhnikov:2006xi,Kusenko:2006rh,Petraki:2007gq,Merle:2013wta,Merle:2015oja,Klasen:2013ypa,Adulpravitchai:2014xna,Shakya:2015xnx,Konig:2016dzg}, are largely unconstrained by present X-ray observations, since they could in principle operate even without active-sterile mixing. Note that the corresponding parent in such Scalar Decay (SD) models must themselves be coupled to the Standard Model, typically via the Higgs sector. Thus, the key free parameters in such settings are the Higgs portal coupling and the Yukawa coupling between the scalar and the sterile neutrinos.\footnote{Again, non-thermal Dark Matter models can suppress so much power at large scales that they are ruled out by the observed cosmic structure. However, the term ``hot Dark Matter'' should be used with care also in this case, since a non-thermal spectrum cannot be associated with a temperature straightforwardly.} For more complicated spectra, they can even look entirely different from thermal cases and exhibit qualitatively new features such as more than one momentum scale, long tales, or similar. 

Another proposed solution to the small-scale problems in galaxy formation is based on Bose condensates of ultra-light (pseudo) scalar field DM with mass $m_\psi\approx 10^{-22}$~eV~\citep{Schive:2014dra,Hu:2000ke,Marsh:2013ywa}, which may be in the form of axions arising in string theory~\citep{Arvanitaki:2009fg} or other extensions of the Standard Model of particle physics~\citep[see][for a historic review]{Kim:1986ax}. In these settings, DM particles can be described by a single coherent wave function with a single free parameter $m_\psi$, the DM particle mass. The quantum pressure arising from the de Broglie wavelength produces a steep suppression of the transfer function below the corresponding Jeans length~\citep{Khlopov:1985jw,Hu:2000ke}, making this scenario a viable alternative solution to the small-scale problems in galaxy formations~\citep[see][]{chavanis2011,Marsh:2013ywa,Woo:2008nn,Mielke:2009zza,Bozek:2014uqa,Martinez-Medina:2014hka,Madarassy:2014jfa,Harko:2015aya}. In fact, the soliton solution allows for cored inner density profiles in dwarf galaxies, while the substructures in DM halos, arising from fine-scale, large-amplitude cellular interference, would yield a suppressed abundance of satellite compared to the CDM case~\citep{du2016}. For such models, often referred to as "Fuzzy" DM, the parameter space corresponding to the different possible power spectra is rather simple, since it depends only on the particle mass $m_\psi$. This is expected to be $m_\psi\lesssim 1.2 \times 10^{-22}$~eV~\citep[see, e.g.,][]{Marsh:2015wka} to resolve the cusp-core problem (without recourse to baryon feedback, or other astrophysical effects), but values up to $m_\psi\approx 4\times10^{-22}$~eV have been considered in previous works~\citep[see][]{Schive:2015kza}. 

The abundance of low-mass cosmic structures provides an important key to either restrict the range of allowed sterile neutrino production models, or rule out presently allowed DM scenarios. In the case of thermal relic WDM particles, the one-to-one correspondence between the WDM particle mass and the suppression in the power spectrum at small scales has allowed to derive limits on $m_X$ by comparing the predictions from $N$-body WDM simulations or semi-analytic models with the abundance of observed ultra-faint satellites. On this basis, different authors have derived limits ranging from $m_X\geq 1.5$~keV~\citep{Lovell:2011rd} to $m_X\geq 1.8$~keV~\citep{Horiuchi:2013noa}, $m_X\geq 2$~keV~\citep{Kennedy:2013uta} and $m_X\geq 2.3$~keV~\citep{Polisensky:2010rw}; relevant constraints have also been obtained for the parameter space of resonant production (RP) sterile neutrino models (Schneider 2016), in that latter case also taking into account the actual shape of the distribution functions (and, thus, not using a WDM approximation to a non-thermal case). Note that, however, such a method is appreciably sensitive to the assumed completeness corrections~\citep[see discussion in][]{Abazajian:2011dt,Schultz:2014eia}, to the treatment of sub-halo stripping, and to the assumed values for the DM mass of the host halo and of the satellites. At higher redshifts, $z\approx 6$, a limit $m_X\gtrsim 1$~keV has been derived from the UV luminosity functions of faint galaxies ($M_{\rm UV}\approx -16$) in~\cite{Schultz:2014eia}; a similar approach by Corasaniti et al. (2016) yield $m_X\gtrsim 1.5$~keV. The same method has been applied to Fuzzy DM models in~\cite{Schive:2015kza} (see also Corasaniti et al. 2016), deriving a consistency of such a model with the observed galaxy abundances for the whole particle mass range, $10^{-22}\leq m_\psi/{\rm eV}\leq 4 \times 10^{-22}$. Since these approaches are based on the comparison between the observed luminosity functions and the predicted mass function of DM halos in different WDM models, the delicate issue in these methods is their dependence on the physics of baryons, determining the mass-to-light ratio of faint galaxies. Although to a lesser extent, uncertainties in the baryonic physics also affect~\citep{Garzilli:2015iwa,Viel:2013apy} the tighter constraints achieved so far $m_X\geq 3.3$~keV for WDM thermal relics, derived by comparing small scale structure in the Lyman-$\alpha$ forest of high- resolution ($z > 4$) quasar spectra with hydrodynamical $N$-body simulations (\cite{Viel:2013apy};  for a generalization of this method to sterile neutrinos models, see Schneider (2016).

To derive robust limits on alternative DM scenarios, it is crucial to bypass the uncertainties related to the physics of baryons involved in galaxy formation, in order to get rid of the degeneracies between the effect of baryons and of the DM power spectrum in suppressing the number of observed low-mass structures. To this aim, \citet{menci2016c,menci2016a} have exploited the downturn of the halo mass distribution  $\phi \left(M,z\right)$ in models with suppressed power spectra, which yields a maximum number density $\overline\phi$ of DM halos in the cumulative mass distribution that in turn depends on the adopted DM model. Since luminous galaxies cannot outnumber DM halos, an observed galaxy density ${\phi}_{obs}>\overline\phi$ would rule out the adopted DM model independently of the baryonic processes determining the luminous properties of galaxies. Such a method, first applied to lensed galaxies at $z=10$ in~\cite{Pacucci:2013jfa} and to galaxies at $z=7$ in the Hubble Deep Field in~\cite{Lapi:2015zea}, has acquired an increased  potential with the first results of the Hubble Frontier Field (HFF) programme. By exploiting the magnification power of gravitational lensing produced by foreground clusters, HFF has enabled the detection of galaxies fainter than the detection limit of the Hubble Deep Field at $z\geq 6$~\citep[see, e.g.][]{Atek:2014kqa,Ishigaki:2014cga,Laporte:2014mca,2016A&A...590A..31C}, to reach unprecedented faint magnitudes $M_{\rm UV}=-12.5$ in the measurement of the luminosity function of galaxies at $z=6$ in~\citep{Livermore:2016mbs}. The large number density ${\phi}_{obs}\geq 1.3\,{\rm Mpc}^{-3}$ (at 2-$\sigma$ confidence level) corresponding to the observed luminosity function allowed us to set a robust lower limit $m_X\geq 2.5$~keV (at 2-$\sigma$) to the mass of thermal relic WDM particles. This constitutes the tightest constraint derived so far on thermal WDM candidates independent of the baryon physics involved in galaxy formation. 

Given the potential and the robustness of this method, it is only natural to apply it to different sterile neutrino production models, which offer a natural framework for physically motivated DM  candidates  as discussed above. In this paper, by requiring the maximum number density attained in a given DM model to be larger than the observed HFF value, we will provide unprecedented constraints on the parameter space for different sterile neutrino production mechanisms. In particular, the combinations $m_{\nu}-\sin^2(2\theta)$ defining the different RP models will be explored to significantly restrict the allowed regions in combination with existing bounds, while for SD models we will set limits on the different combinations of free parameters, i.e., the scalar mass, the Higgs portal coupling, and the Yukawa coupling between the scalar and the sterile neutrino. We shall then turn to apply the method to the Fuzzy DM scenario to provide lower limits on the ultra-light pseudo-axion mass $m_\psi$ which will turn out to rule out this class of models. 

The paper is organized as follows. In Sect.~\ref{sec:Method} we provide a description of the method. In Sect.~\ref{sec:DMModels} we briefly summarize the basic features of the DM models we are probing in this work, i.e., the RP and the SD production models for sterile neutrinos, and the Fuzzy DM model. In Sect.~\ref{sec:Results} we present and discuss the results, while Sect.~\ref{sec:Conclusions} is devoted to our conclusions. 


\section{\label{sec:Method}Method}

To derive constraints on the parameter space of sterile neutrino (RP and SD) DM models and of the Fuzzy DM models, we compute the maximum number density of DM halos $\overline{\phi}$ expected at redshift $z=6$ for each point of the parameter space, and compare it to the observed number density ${\phi}_{obs}$ of galaxies at the same redshift obtained from the galaxy luminosity function measured by \cite{Livermore:2016mbs}. Since observed galaxies cannot outnumber their host DM halos, the condition $\overline{\phi}\geq {\phi}_{obs}$ determines the set of parameters admitted for each DM model. 

The method is based on the drop of the differential halo mass function $\diffd \phi/\diffd M$ at small masses in models where the power spectrum is strongly suppressed with respect to CDM at masses $M\lesssim 10^7-10^9\,M_\odot$. As a consequence, the corresponding cumulative halo mass function $\int^{\infty}_M\,\diffd m\; ({\diffd \phi/ \diffd m})$ saturates to a maximum value $\overline{\phi}$ when the integral is extended down to progressively smaller values of $M$. This provides a maximum value for the number density of DM halos regardless of the underlying mass-luminosity relation, and -- hence -- \emph{completely independent of the baryon physics entering galaxy formation}. 

In the following, we describe our computation of the observed number density  ${\phi}_{obs}$ and or the maximum predicted number density $\overline{\phi}$, in turn. 

\subsection{The galaxy number density at $z=6$ from observed luminosity functions}

For the observed number density ${\phi}_{obs}$, we take the value derived integrating the galaxy luminosity function at $z=6$ down to the faintest bin $M_{\rm UV}= -12.5$~\citep{Livermore:2016mbs}. 
The luminosity function  has been estimated from objects in the Abell~2744 and MACS~0416 cluster fields, selected on the basis of their photometric redshift. The effective volume sampled by the observations was computed from simulations of $5\times10^5$ mock galaxies with realistic colors and shapes, using all lensing models available for the two fields in the MAST archive. The simulated sources are added to the images and analysed with the same procedure used for real sources, including photometric redshift analysis, in order to estimate the flux-dependent selection completeness. The resulting fiducial luminosity function with the corresponding 1-$\sigma$ uncertainties in each magnitudebin) is estimated on the basis of the median magnification for each galaxy in the sample and is reported in Fig.~10 of~\citet{Livermore:2016mbs}. 
 From this we have derived the observed cumulative number density $\phi_{obs}$ (and its 
confidence levels) through a Monte Carlo procedure \citep{menci2016c}. We extracted 
random  values $\Phi_{random}(M_{UV})$ of the luminosity function in each magnitude bin according to a Gaussian distribution with 
variance given by the error bar in \cite{Livermore:2016mbs}.  Thus, for each simulation we produced a new realization of the luminosity function at $z=6$. From this, a
cumulative number density $\phi_{random}$ has been derived by summing up the values of $\Phi_{random}(M_{UV})$ in all the
observed magnitude bins in the range  $-22.5\leq M_{UV}\leq -12.5$.  We carried out $N_{sim}=10^7$ simulations to compute the probability
distribution function (PDF) of the cumulative number density $\phi_{random}$. We obtain a median value  $log\,{\phi}_{obs}/{\rm Mpc}^{-3}=0.54$, 
 while from the relevant percentiles of the PDF we  derive lower bounds  0.26, 0.01, and -0.32 at  1, 2, and 3-$\sigma$ confidence levels, respectively. We have checked that the median value of the differential luminosity function $\Phi_{random}$ obtained from our simulations  
is consistent (within 3\%) with the best fit value of the  luminosity function obtained by \cite{Livermore:2016mbs}.   Note that such a procedure assumes that  each magnitude bin is uncorrelated with the adjacent ones (as indeed done in both the Livermore 2016 and the Bouwens 2016 analysis). 
 In case of correlated bins larger errorbars are expected for 
blank field observations (see the discussion in Castellano et al. 2010), but a quantitative estimate of such an effect for measurements involving lensing magnification is lacking; this would constitute an interesting improvement over the present treatment. 

 Note that the measurements of the luminosity functions derived by \cite{Livermore:2016mbs}  are particulary delicate at the faint end where large lensing magnifications are involved. Indeed, Bouwens et al. (2016) have adopted 
a different estimate of the impact of lensing magnifiction finding not only a lower median value for the number density of galaxies 
at $M_{UV}=-12.5$ compared to \cite{Livermore:2016mbs}, but also  larger errorbars. In fact, assuming the luminosity functions in 
Bouwens et al. (2016) we find for the maximum number density a median value  $\log {\phi}_{obs}/{\rm Mpc}^{-3}=-0.25$, with lower bounds  $-0.47$, $-0.62$, and $-0.9$ (at  1, 2, and 3-$\sigma$ confidence levels), yielding looser limits on the parameter space of DM models. We  discuss the impact of assuming such values for $\phi_{obs}$ in Sect.~\ref{sec:Conclusions}. In the same section we present a critical discussion of the systematics associated to the  measurements of  highly magnified sources.

\subsection{The maximum number density of halos in dark matter models with suppressed power spectra}

The computation of the differential halo mass function ${\diffd\,\phi/ \diffd(\log M)}$ in sterile neutrino models is based on the standard procedure described and tested against $N$-body simulations in~\cite{Schneider:2011yu,Schneider:2013ria,Benson:2012su,Angulo:2013sza}. Our computation has been tested against simulations in~\cite{menci2016a} and presented also in~\cite{menci2016c}. Here we summarize the key points of the computation and we refer to the above papers for details. 

The key quantity entering the computations is the variance of the linear power spectrum $P(k)$ of DM perturbations (in terms of the wave-number $k=2\pi/r$). Its dependence on the spatial scale $r$ of perturbations is given by
\begin{equation} 
{\diffd \; (\ln\,\sigma^2) \over \diffd \; (\ln\,r)}=-{1\over 2\,\pi^2\,\sigma^2(r)}\,{P(1/r)\over r^3}.
\label{eq:VarianceDerivative}
\end{equation}
Here we have used a sharp-$k$ form (i.e., a top-hat sphere in Fourier space) for the window function $W(kr)$ relating the variance to the power spectrum $\sigma^2(M)=\int dk\,k^2\,P(k)\,W(kr)/\left(2\,\pi^2\right)$ -- see e.g. \cite{Schneider:2013ria} for other common choices. The normalization $c$ entering the relation between the halo mass $M = 4\pi\,\overline{\rho} (cr)^3/3$ and the filter scale $r$ must be calibrated through simulations (here, $\overline{\rho}$ is the background density of the Universe).  All studies in the literature yield values for $c$  in the range $c = 2.5-2.7$~\citep[see, e.g.,][]{Angulo:2013sza,Benson:2012su,Schneider:2013ria}. The effect of such an uncertainty will be considered in deriving the constraints presented in Sect. 4.

For the sterile neutrino RP and SD models, the power spectrum is computed directly by solving the Boltzmann equation after computing the distribution function for all points of the parameter space, as described in detail in Sect.~\ref{sec:DMModels}. Then, the differential halo mass function (per unit $\log\,M$) based on the extended Press \& Schechter approach~\citep{Bardeen:1985tr,Benson:2012su,Schneider:2013ria} reads
\begin{equation}
 {\diffd \,\phi \over \diffd \left(\ln M\right)}={1\over 6}\,{\overline{\rho}\over M}\,f(\nu)\,{\diffd \left(\ln\,\sigma^2\right) \over \diffd \left(\ln \; r\right)}\,.
 \label{eq:HaloMassDerivative}
\end{equation}
Here, $\nu\equiv \delta_c^2(t)/\sigma^2$ depends on the linearly extrapolated density for collapse in the spherical model, $\delta_c=1.686/D(t)$, and $D(t)$ is the growth factor of DM perturbations. We conservatively assume a spherical collapse model, for which $f(\nu)=\sqrt{2\nu/\pi}\,\exp(-\nu/2)$. Assuming an ellipsoidal collapse model would yield a lower halo mass function at the low-mass end and -- hence -- even tighter constraints on the DM particle mass.

For sterile neutrino DM, yielding power spectra suppressed at small scales compared to CDM, the resulting differential mass functions, see \equref{eq:HaloMassDerivative}, are characterized by a maximum value at masses close to the ``half-mode'' mass~\citep{Schneider:2011yu,Benson:2012su,Schneider:2013ria,Angulo:2013sza},  the mass scale at which the spectrum is suppressed by 1/2 compared to CDM. This function depends strongly on the sterile neutrino mass; for RP models it also depends on the lepton asymmetry assumed and, hence, on the resulting mixing angle $\theta$; typical power spectra in such models yield half-mode masses ranging from $M_{hm}\approx 10^{10}\,M_\odot$ to $M_{hm}\approx 10^{8}\,M_\odot$. Correspondingly, the cumulative mass functions saturate to a maximum value  $\overline{\phi}(z)\approx \phi(M_{hm},z)$, defining the maximum number density of DM halos associated to the considered power spectrum. 

A similar behavior characterizes the halo mass functions in the Fuzzy DM case. Dedicated $N$-body simulations~\citep{Schive:2015kza} yield for the differential mass function the form 
\begin{equation}
 {\diffd\,\phi\over \diffd (\ln M)}={\diffd \,\phi\over \diffd (\ln M)}\bigg|_{\rm CDM} \cdot \Bigg[1+\bigg({M\over M_0}\bigg)^{-1.1}\Bigg]^{-1.2},
  \label{eq:HaloMassDerivativeFuzzy}
 \end{equation}
where $\left |\diffd \,\phi/ \diffd \left(\ln M\right)\right|_{\rm CDM}$ is the halo mass function in the CDM scenario computed after Eqs. 1 and 2, assuming for $P(k)$ the CDM form given in~\cite{Bardeen:1985tr}. The auxiliary mass scale $M_0 = 1.6 \times 10^{10}\, (m_\psi/10^{-22}\,{\rm~eV})^{-4/3}\, M_\odot$, determining the  suppression of the halo mass function compared to the CDM case, depends on the Fuzzy DM candidate mass, and it plays a role analogous to the half-mode mass scale for sterile neutrino models.\footnote{Instead of using the fitting function given by \cite{Schive:2015kza}, one could also apply \equref{eq:HaloMassDerivative} to the case of Fuzzy DM. We use \equref{eq:HaloMassDerivativeFuzzy} for simplicity and because this guarantees full consistence with previous work on Fuzzy DM.} For $m_\psi\geq 1.5 \times 10^{-22}\,{\rm~eV}$ the uncertainties in the above expression for the halo mass functions are below 10\%~\citep{Schive:2015kza} and will be considered when comparing with observed galaxy number densities. 

For each DM model considered, the method described above allows to exclude the region of the parameter space yielding ${\phi}_{obs}>\overline{\phi}$. In the next Section, we briefly recall the properties of the parameter space we explore for the different DM models we consider.


\section{\label{sec:DMModels}The Dark Matter Models: Power Spectra and the Parameter Space}

\subsection{\label{sec:DMModels:SFProd}Sterile neutrinos: resonant production from mixing with active neutrinos}

The minimal setup for sterile neutrino DM is the production via mixing with one or several active neutrino flavors. Active neutrinos are weakly interacting and are therefore in thermal equilibrium with other Standard Model particles in the early Universe (i.e.~at temperatures above the MeV range). During that epoch, the sterile neutrino abundance builds up gradually via occasional oscillations from the active to the sterile sector. Depending on mixing angle and particle mass, this \emph{freeze-in} production can lead to the right amount of sterile neutrino DM~\citep{Dodelson:1993je} and is usually referred to as Dodelson-Widrow (DW) mechanism or \emph{non-resonant} production.\footnote{Note that the original proposal of this mechanism was prior to DW in \cite{Langacker:1989sv}, however, DW were the first to link it to DM.} Recent investigations based on combined limits from structure formation and X-ray observations have ruled out this production mechanism as the dominant contributor to the DM sector~\citep{Seljak2006,Viel2006,Horiuchi:2013noa,Merle:2015vzu}. 

However, it has been noticed early on that the active-sterile oscillation may be enhanced by a resonance~\citep{Shi:1998km}, provided there exists a significant lepton asymmetry $L$ in the early Universe. Such a resonance allows for significantly smaller mixing angles $\theta$, relaxing the tight limits from X-ray observations. Furthermore, the resulting particle momentum distributions may be colder and, therefore, in better agreement with structure formation~\citep{Abazajian:2001nj}. The Shi \& Fuller (SF) or \emph{resonant} production mechanism of sterile neutrino DM~\citep{Shi:1998km}, first mentioned in \cite{Enqvist:1990ek}, also plays an important part in the framework of the \emph{Neutrino Minimal Standard Model} ($\nu$MSM), which attempts to simultaneously solve the problems of DM, non-zero neutrino masses, and baryon asymmetry by solely adding three additional sterile neutrino flavors to the Standard Model~\citep[see e.g.][]{Asaka:2005an,Asaka:2005pn,Canetti:2012kh}, at the cost of introducing a considerable fine tuning to produce a suitable lepton asymmetry. Note that in the $\nu$MSM, there is an upper limit for the allowed lepton asymmetry \citep[see][]{Canetti:2012kh,Shaposhnikov:2008pf,Laine:2008pg} and hence, there is a lower limit on the mixing angle even in the resonant case if the production mechanism is to explain the total DM abundance.

Since, for any given sterile neutrino mass, the mixing angle is related to the adopted lepton asymmetry $L$, in this work we describe the parameter space of RP sterile neutrino models in terms of combinations of sterile neutrino masses $m_{\nu}$ and mixing amplitudes $sin^2(2\theta)$. Each one of such combinations corresponds to a different momentum distribution, which strongly differs from a generic Fermi-Dirac form~\citep{Abazajian:2001nj}. In many cases such  non-thermal momentum distributions are characterized by colder mean particle momenta and by a larger  range of different momenta compared to the case of non-resonant production. In terms of structure formation, this may result in density perturbations which are more gradually suppressed and survive down to smaller scales. This is, however, only true for certain parts of the parameter space, while other parts show similar (or even stronger) suppression of matter perturbations compared to the case of non-resonant sterile neutrino DM (for more details see Schneider 2016).

In this work the sterile neutrino momentum distributions of RP are computed with the public code {\it sterile-dm} of~\cite{Venumadhav:2015pla}, which is an extension of earlier works \citep{Abazajian:2001nj,Kishimoto:2008ic,Abazajian:2014gza}. The computation is based on the Boltzmann equation and includes detailed calculations of the lepton asymmetry around the quark-hadron transition. Independent calculations by \cite{Ghiglieri:2015jua} give similar results. To obtain the power spectra, we use the publicly available Boltzmann solver {\it CLASS}~\citep{Blas:2011rf,Lesgourgues:2011rh}.

\subsection{\label{sec:DMModels:SDProd}Sterile neutrinos: production from scalar decay}

Production from scalar decay is described by a generic model that invokes one real scalar singlet $S$ and (at least) one sterile neutrino $N$ beyond the Standard Model. The scalar singlet couples to the SM Higgs doublet $\Phi$ via a \emph{Higgs portal},
\begin{equation}
 \mathcal{L} \supset 2\lambda \left(\Phi^\dagger \Phi \right) S^2 \,,
 \label{eq:HiggsPortalLag}
\end{equation}
where $\lambda$ is a dimensionless coupling. The interaction between the scalar and the sterile neutrino is encoded in
\begin{equation}
 \mathcal{L} \supset - \frac{y}{2} S\overline{N^c}N \,,
 \label{eq:YukawaLag}
\end{equation}
where $y$ is a Yukawa-type coupling. In the most general case, the complete model Lagrangian should contain terms for the mixing between active and sterile neutrinos. We will, however, neglect these terms, since non-resonant active-sterile mixing cannot contribute significantly to the production of sterile neutrinos when X-ray limits and limits of structure formation are taken into account~\citep{Merle:2015vzu}. Note that the assumption of having only one right-handed neutrino is not very restrictive, since most statements can easily be generalised to the case of several right handed states~\citep{Konig:2016dzg}.

The free parameters of the scalar decay model are the following:
\begin{enumerate}
 \item The Higgs portal coupling $\lambda$, which determines the production rate and the kinematics of the scalar from the SM degrees of freedom of the Higgs doublet. 
 \item The Yukawa coupling $y$, which enters into the decay rate of the scalar and hence controls how fast the scalar decays into sterile neutrinos.
 \item The mass of the scalar singlet, $m_S$, which determines which channels contribute to the production of scalars and thereby finally to the abundance of sterile neutrinos. If $m_S < m_h/2$, where $m_h$ is the mass of the physical Higgs boson, thermal Higgses can directly decay into scalars which increases the production drastically compared to those cases where $S$ can only be produced from pairwise annihilation of SM degrees of freedom. For a detailed discussion of the individual regimes, see Sect. 2 of ~\cite{Konig:2016dzg}.
 \item The mass of the sterile neutrino, denoted $m_\nu$. It will strongly influence the effects on cosmological structure formation. Nonetheless we want to stress again that the scalar decay mechanism is highly non-thermal and that no premature conclusion on the sterile neutrino being ``too hot'' or ``quasi-cold'' should be made.
\end{enumerate}
In this work, we will treat $\lambda$, $y$, and $m_S$ as free parameters. For each triple of $\left(\lambda, y, m_S\right)$, we fix the mass of the sterile neutrino by requiring it to reproduce the observed relic DM abundance. We use the best-fit values published by the Planck collaboration~\citep{Planck:2015xua}. We have scanned the remaining three-dimensional parameter space by first solving the homogeneous and isotropic Boltzmann equation governing the production of sterile neutrinos in the early Universe~\citep{Konig:2016dzg}. The resulting momentum distribution functions and the value of $m_\nu$ inferred from the relic abundance constraint then served as an input to compute the linear power spectra, again using the {\it CLASS} code~\citep{Blas:2011rf,Lesgourgues:2011rh}.

Let us very briefly discuss the interplay between the Higgs portal and the Yukawa coupling, in order to make the results presented later on easier to digest: for small Higgs portal couplings, the scalar itself is produced by freeze-in and is always strongly suppressed compared to its would-be equilibrium abundance. Physically, this means that backreactions of scalars into SM particles can be neglected completely. In this case, the relic abundance of sterile neutrinos (and hence the mass $m_\nu$) are independent of the Yukawa coupling $y$ for a fixed pair  $\left(m_S,\lambda\right)$. Nonetheless, the Yukawa coupling controls the production time of the sterile neutrinos, which is one of the key factors for structure formation. When $\lambda$ is large enough to equilibrate the scalars, they will be subject to the well-known dynamics of freeze-out. In this regime, sterile neutrinos can be produced from scalars in equilibrium and from those decaying after freeze-out. Accordingly, the number density of steriles and thereby their mass $m_\nu$ can strongly depend on $y$ even for fixed $\left(m_S,\lambda\right)$. Again, we refer the reader 
to \cite{Konig:2016dzg} for a more detailed discussion.

\subsection{\label{sec:DMModels:WaveDM}Fuzzy Dark Matter}

Fuzzy DM models assume the DM to be composed of a non-relativistic Bose-Einstein condensate, so that the uncertainty principle counters gravity below a Jeans scale corresponding to the de Broglie wavelength of the ground state. In this case, the  suppression of  small scale structures and the formation of galactic cores in dwarf galaxies is in fact entirely due to the uncertainty principle, which counteracts gravity below the Jeans scale, corresponding to a mass scale $M_J=10^{7}\,M_{\odot}\,m_{22}^{-3/2}$~\citep{Marsh:2013ywa}, where $m_{22}\equiv m_{\psi}/10^{-22}$ eV. In such models, the DM mass $m_{\psi}$ ultimately determines all the relevant DM physical scales in structure formation, since it determines the scale below which an increase in momentum opposes any attempt to confine the particle any further. E.g., the inner halo density profiles in such models are well described by the stable soliton solution of the Schr\"odinger-Poisson equation, which extends up to a core radius of $r_{\psi}=1.6\,m_{22}^{-1}\,(M/10^9\,M_{\odot})^{-1/3}$~kpc~\cite{Schive:2014dra};  at larger radii, the properties of Fuzzy DM halos are indistinguishable from CDM. Recent works comparing the observed stellar-kinematical data of dwarf spheroidal galaxies to the density profiles produced in Fuzzy DM scenarios derived upper limits for $r_{\psi}$ which translate in constraints on the DM particle mass  $m_{\psi}\lesssim 1.5\cdot10^{-22}$~\citep{Marsh:2015wka,calabrese2016}.
 
Since the DM particle mass is the only free parameter in Fuzzy DM models, comparing the mass distribution of collapsed DM halos derived by~\citep{Schive:2015kza}, Eq.~\eqref{eq:HaloMassDerivativeFuzzy}, with the observed abundance of high redshift galaxies as described in Sect. 2,  provides straightforward constraints on $m_{\psi}$ which can be compared with the existing bounds mentioned above.


\section{\label{sec:Results}Results}

We now proceed to present the constraints on the parameter space of the different models presented above. In all cases we adopted the Planck values for the cosmological parameters~\citep{Planck:2015xua}.
\subsection{\label{sec:Results:SFProd}Resonant production of sterile neutrinos}

In the case of resonantly produced sterile neutrino DM, we choose the free parameters to be the mass, $m_{\nu}$, and the mixing amplitude $\sin^2(2\theta)$. For each combination of such quantities, the lepton number $L$ is fixed to the value required to yield the right DM abundance. We first investigate the effect of varying the mixing angle for a fixed sterile neutrino mass by focusing on the case $m_{\nu}=7.1$~keV, corresponding to a sterile neutrino whose decay could be at the origin of the potential 3.5~keV line in X-ray spectra of clusters. For such a case, the spectra computed as presented in Sect.~\ref{sec:DMModels} yield the cumulative halo mass functions  shown in Fig. 1 (left panel) for different values of $\sin^2(2\theta)$ and compared with the observed number density of galaxies with $M_{\rm UV}\leq -12.5$ in the HFF. Note that the lines in Fig. 1 constitute  upper limits with respect to the theoretical uncertainties discussed in Sect.~\ref{sec:Method}. At small masses, they saturate to a maximum number density $\overline{\phi}$, which is plotted in the right panel of Fig. 1 as a function of $\sin^2(2\theta)$. When compared with the observed number density of luminous galaxies from the HFF (the upper shaded areas), requiring the number density of DM halos to be larger than the observed abundance $\overline{\phi}\geq {\phi}_{obs}$ restricts the mixing angle in the range $2\times10^{-11}\leq \sin^2(2\theta)\leq 10^{-9}$ (at 2-$\sigma$ confidence level). When combined with existing upper bounds from X-ray observations \citep[hatched region in the right-hand panel of Fig.1; see also][]{Watson:2011dw,Horiuchi:2013noa}, this restricts the range of allowed values to the small interval $2\times10^{-11}\leq \sin^2(2\theta)\leq 10^{-10}$. 

\begin{center}
\hspace{-0.cm}
  \includegraphics[width=0.46\textwidth]{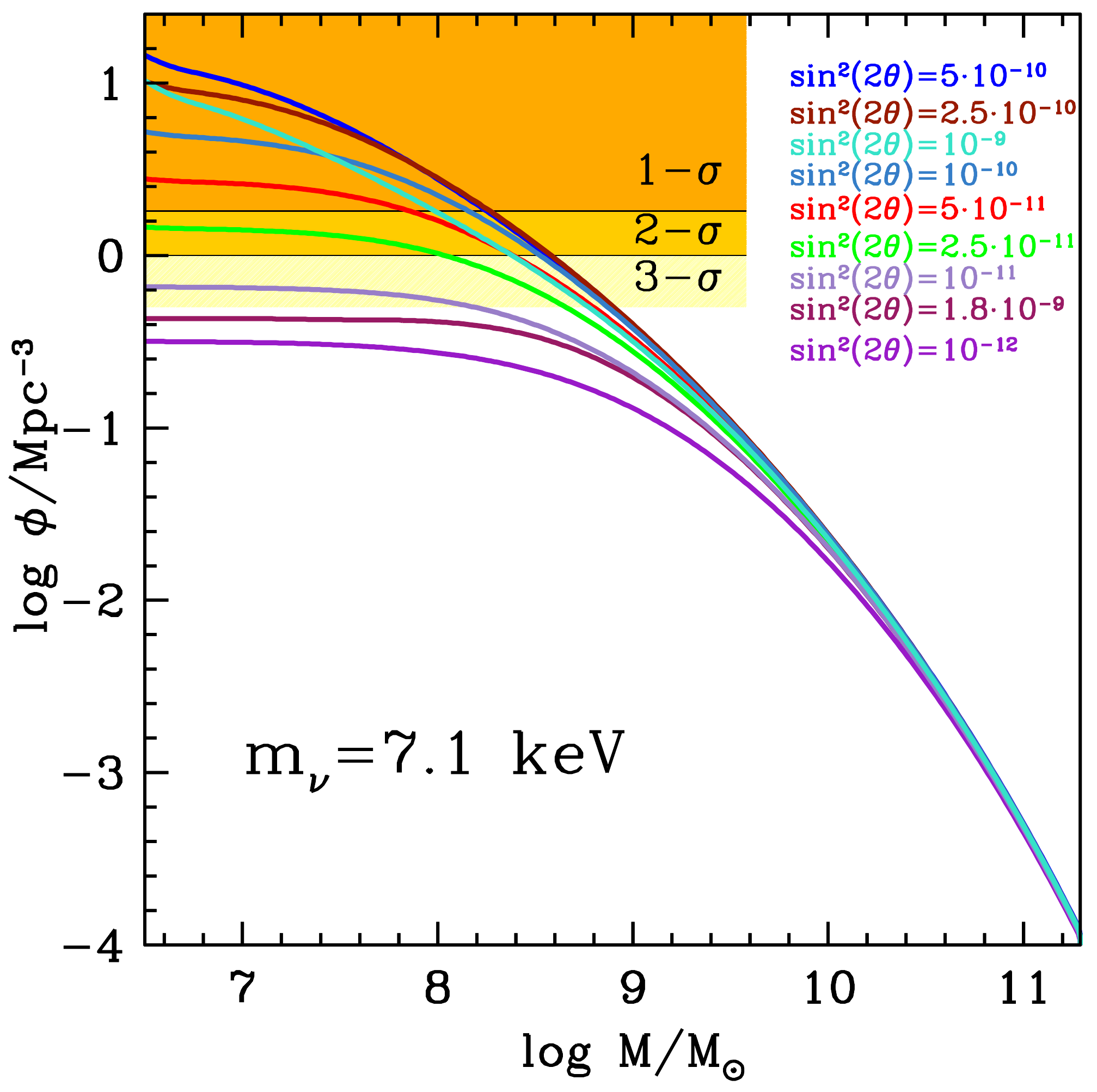}
  \includegraphics[width=0.52\textwidth]{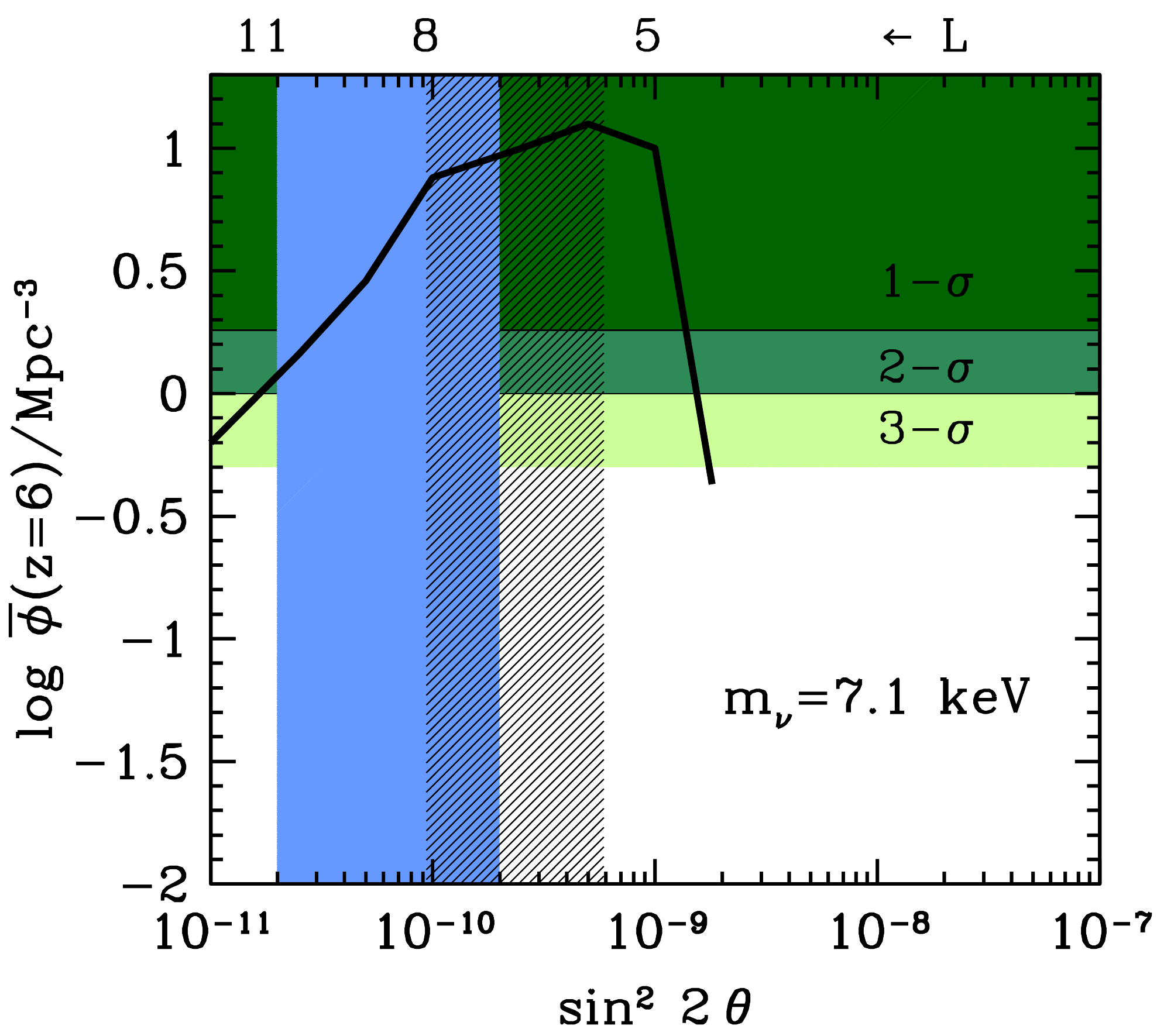}
  \end{center}
  \vspace{-0.5cm}
{\footnotesize Fig. 1. Left Panel: The cumulative mass functions computed at $z=6$ for RP sterile neutrino models with $m_{\nu}=7.1$~keV for different values of the mixing angle $\theta$ shown by the labels on the right. The shaded areas correspond to the observed number density of galaxies with $M_{\rm UV}\leq -12.5$ within within 1-$\sigma$, 2-$\sigma$, and  3-$\sigma$ confidence levels. Right Panel: The maximum value $\overline{\phi}$ (including theoretical uncertainties) of the predicted number density of  DM halos at $z=6$ for the case with $m_{\nu}=7.1$~keV as a function of the mixing angle $\theta$. The upper shaded areas represent the observed number densities of galaxies with $M_{\rm UV}\leq -12.5$ within the 1-$\sigma$, 2-$\sigma$, and  3-$\sigma$  confidence levels. The vertical filled area corresponds to the range of values of $\sin^2(2\theta)$ consistent with the tentative line signal 3.5~keV in X-ray spectra~\citep{Bulbul:2014sua,Boyarsky:2014jta}, with the hatched vertical area corresponding to present upper limits on $\sin^2(2\theta)$  from the absence of such a line in the spectra of the Milky Way and of dwarf galaxies from several authors, as given in~\cite{Riemer-Sorensen:2014yda} and references therein.}
\vspace{0.2cm}

To constrain different combinations of the free parameters and to compare with previous results, we explore the whole range of free parameters using a grid of values for both $m_{\nu}$ and $\sin^2(2\theta)$. After computing the corresponding power spectra (Sect.~\ref{sec:DMModels}), the condition $\overline{\phi}\geq {\phi}_{obs}$ leads to the exclusion region in the plane $m_{\nu}-\sin^2(2\theta)$ shown in Fig. 2. 

Note that the exclusion plot is characterized by two  excluded regions. The upper region is bounded by the non-resonant DW limit $L=0$ (the upper green curve in Fig. 2) and is characterized by non-thermal spectra with resonant peaks at low momenta in the momentum distribution. This leads to a strong suppression of the power spectrum yielding the upper exclusion region. For lower values of $\sin^2(2\theta)$, the corresponding increase of the lepton asymmetry (see Sect.~\ref{sec:DMModels:SFProd}) shifts the amplitude and position of the resonance peaks toward larger momenta so that the overall spectrum becomes cooler. The corresponding predicted number density of galaxies $\overline{\phi}$ becomes large enough to be consistent with the observed values $\phi_{\rm obs}$, and results into the allowed region of Fig. 2. When the lepton number is further increases, i.e., for even smaller values of $\sin^2(2\theta)$ the resonance peak in the momentum distribution shifts to sufficiently high momenta to make the spectrum warmer again, thus yielding the lower exclusion region. 
 
\begin{center}
 \includegraphics[width=0.78 \textwidth]{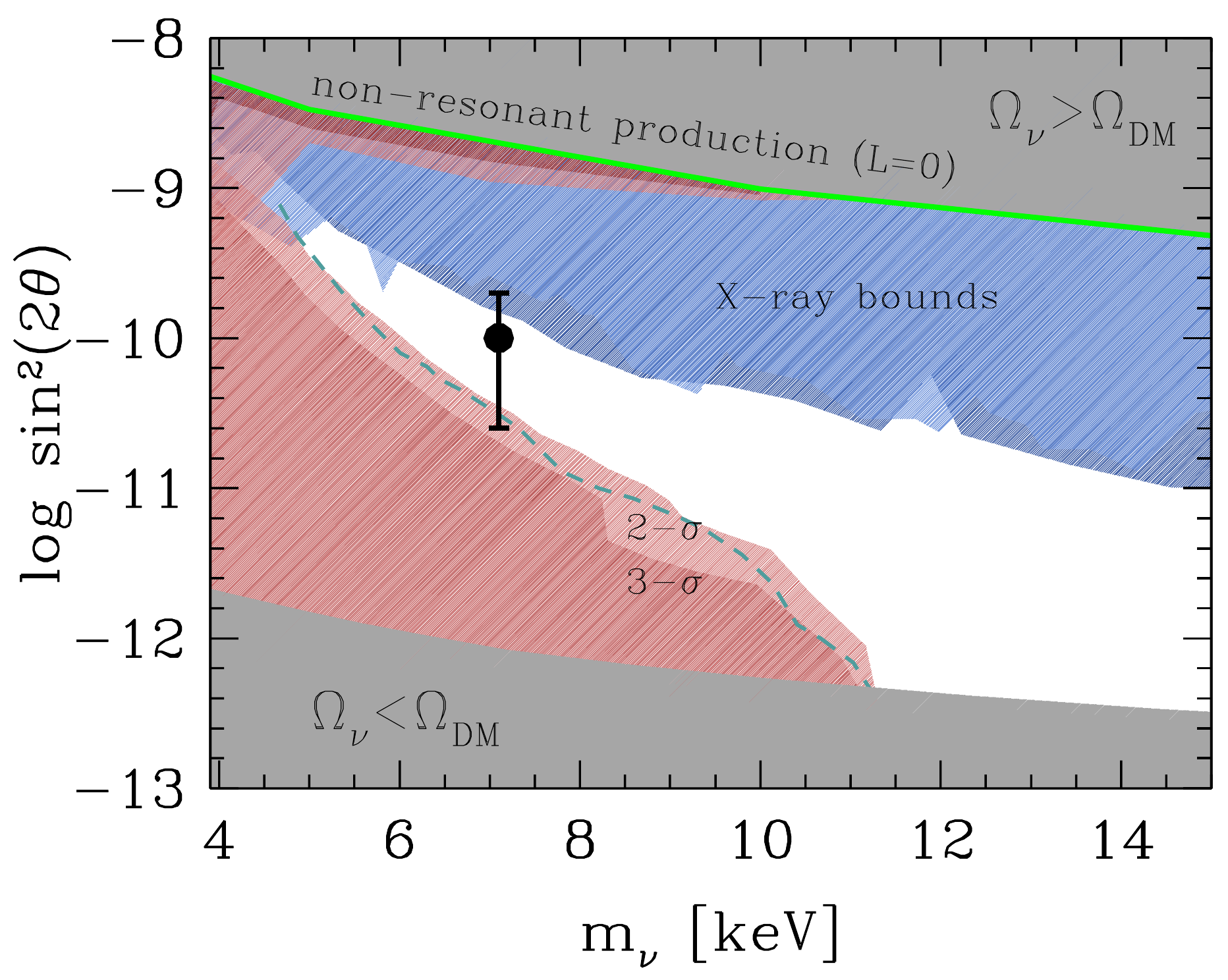}
\end{center}
\vspace{-0.5cm}
{\footnotesize  Fig. 2. The constraints on the RP sterile neutrino parameter space from our method are represented as exclusion regions,    with 3-$\sigma$ and 2-$\sigma$ limits represented by darker and lighter colors. Our constraints are compared with upper bounds from X-ray observations~\citep[as reported in][]{Riemer-Sorensen:2014yda} of the Milky Way and dwarf galaxies. Grey areas are excluded by current limits on the abundance of DM~\citep{Planck:2015xua}. The green line  corresponds to the non-resonant DW case with vanishing lepton asymmetry $L=0$. We also show as a dashed line the constraint obtained in Schneider (2016) from the abundance of satellites in the Milky Way. The tentative line signal at 7.1~keV~\citep{Bulbul:2014sua,Boyarsky:2014jta} is shown by the point with errorbars.  }
\vspace{0.2cm}

We exclude all models with a sterile neutrino mass below $m_{\nu}\leq 5$~keV and also large parts of the parameter space above. The region around the 3.5~keV line ($m_{\nu}=7.1~keV$) is still allowed, but restricted to a narrow range of values for $-11.3\leq\log \sin^2(2\theta)\leq-10$. Our lower bounds on the mixing angle extend over a wide range of sterile neutrino masses up to $m_{\nu}\approx 11$~keV, where the mixing angle is constrained to the interval $-12\leq\log \sin^2(2\theta)\leq-11.4$. Our results constitute the tightest lower bounds on the mixing angle (and hence on the lepton asymmetry) derived so far by methods independent of baryonic physics. For a large range of neutrino masses, our results are very close to the limits obtained in~\cite{Schneider:2016uqi} from the local abundance of Milky Way satellites, using a counting based on the extended Press \& Schechter approach, calibrated to match the total abundance of satellites measured in test cases through $N$-body simulations. The similarity of these results is encouraging, given the different method adopted and, most of all,  the large difference in cosmic times between the observations used to constrain the models. However, it must be noted that estimates of the Milky Way satellites depend on the assumed upper limit for the DM halo of the Milky Way ($3\cdot10^{12}\,M_\odot$), on the assumed isotropic distribution of satellites, on the treatment of sub-halo stripping, and on the assumed lower limit for the DM mass satellites \citep[$10^{8}\,M_\odot$; for more details on the approach, see][]{Schneider:2014rda}. \citet{Lovell:2015psz} adopted a similar approach based on satellite abundances computed through a semi-analytic computation and found a somewhat looser limits, leaving $\sin^2\left(2\theta\right)$ unconstrained above $m_{\nu}\sim 8$~keV. 

On the other hand, our results are less stringent compared to the method based on the small scale structure in the Lyman-$\alpha$ forest of high-resolution ($z > 4$) quasar spectra with hydrodynamical $N$-body simulations~\citep{Viel:2013apy}. This method has been directly applied to warm, thermal DM (yielding a lower limit $m_X\geq 3.3$~keV), and later extended to sterile neutrinos (Schneider 2016) by considering the values and the slopes of the suppression of the one-dimensional power spectrum consistent with the observed Lyman-$\alpha$ structures. However, such a method depends on baryonic physics, and in particular on the thermal history of the intergalactic medium. Indeed, when assumptions concerning this quantity are relaxed, the Lyman-$\alpha$ limit on the thermal relic mass is reduced by about 1 keV~\citep{Viel:2013apy,Garzilli:2015iwa}, and becomes comparable to the constraint from the high-redshift galaxies~\citep[see][]{menci2016c}.

\subsection{\label{sec:Results:SDProd}Scalar decay production of sterile neutrinos}

As discussed in detail in Sect.~\ref{sec:DMModels}, in this case the parameter space is three-dimensional, since it includes the mass of the scalar $m_S$, the Higgs portal coupling $\lambda$ and the Yukawa coupling with the scalar $y$. The sterile neutrino mass $m_{\nu}$ is related to a combination of the three free parameters by demanding that the model reproduces the observed relic abundance of dark matter. Before performing a complete exploration of the parameter space, we first show  a comparison between the model cumulative halo distributions and the observed number density of galaxies in the illustrative case of a sterile neutrino with $m_{\nu}=7.1$~keV (the candidate origin of the potential 3.5~keV line) in the limit of small Higgs portal coupling $\lambda\ll 10^{-6}$. In such a freeze-in limit (see Sect.~\ref{sec:DMModels}), the scalar mass $m_S$ and the Yukawa coupling $y$ will be related for any given value of $m_{\nu}$, thus reducing the parameter space to be explored. In Fig. 3 (left panel) we show the cumulative halo distributions in such a regime for a scalar mass of $m_S = 60$~GeV and for different values of the Yukawa coupling $y$, and we compare them to the observed number density of high-redshift HFF galaxies. The strong dependence of the maximum predicted number density $\overline{\phi}$ on $y$ is highlighted in the right-hand panel of Fig. 3, which shows that present data allow to set a constraint $y\geq 9\times10^{-9}$ at 2-$\sigma$ confidence level. 

We then extend our exploration to cover the whole parameter space of SD production model for sterile neutrinos. To this aim, we consider a grid of $\lambda$ and $y$ values for six different values of the scalar mass $m_S/{\rm~GeV}=60, 65, 100, 170, 500, 1000$. The latter are chosen as to cover the three different regimes where different production channels are dominant for the production of scalars from their coupling to the SM particles \citep[see][and Sect.~\ref{sec:DMModels}]{Konig:2016dzg}: the light scalar case $m_S\leq m_h/2$, the intermediate case $m_h/2\leq m_S\leq m_h$, and the heavy scalar case $m_S\geq m_h$ (in terms of the Higgs mass $m_h=125$~GeV). These different regimes have a rather strong impact on the distribution function of the scalar which  directly translates into that of the sterile neutrino and, hence, on the resulting power spectrum. 
 
\begin{center}
  \includegraphics[width=0.49\textwidth]{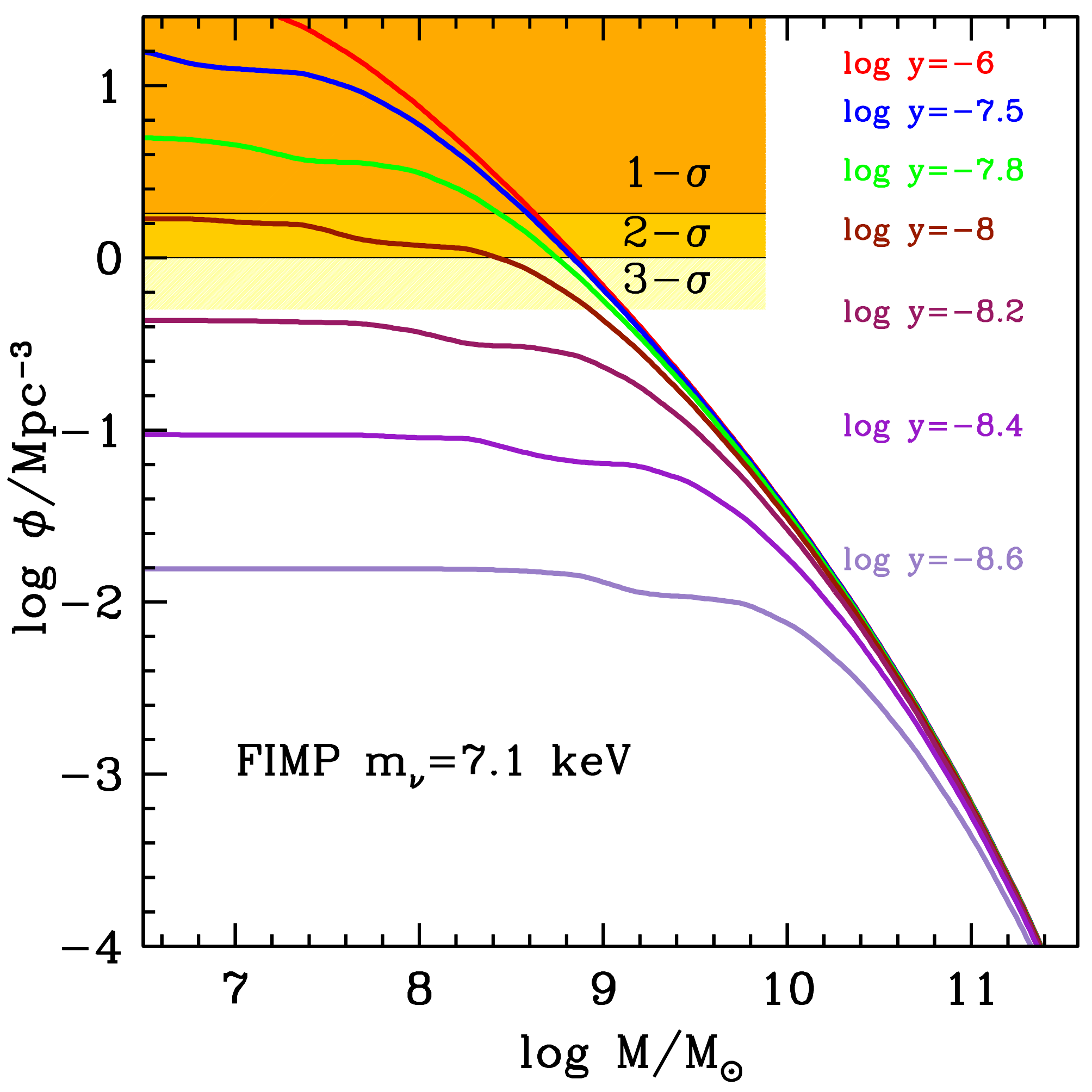}
  \includegraphics[width=0.49\textwidth]{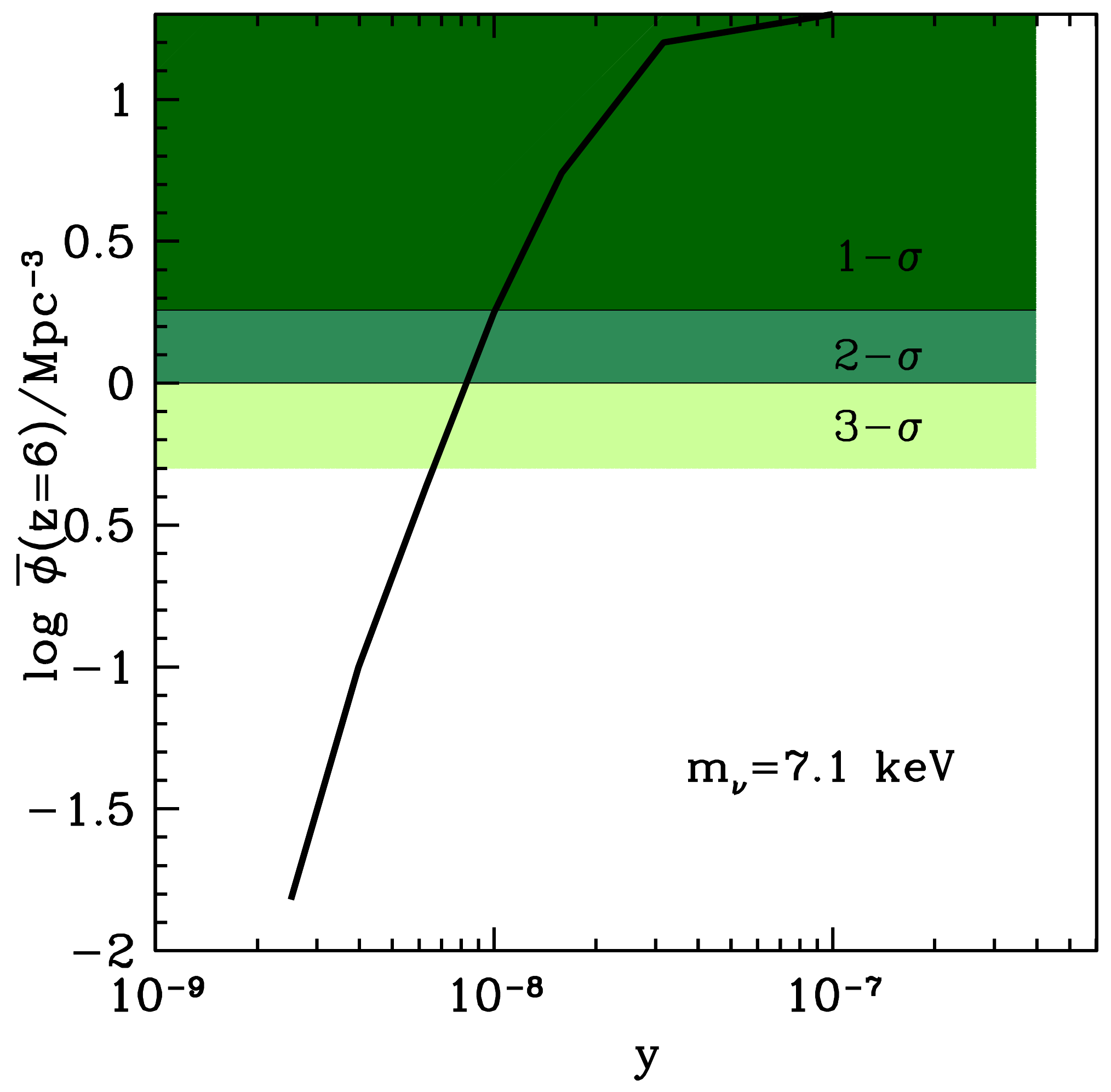}
\end{center}
\vspace{-0.5cm}
{\footnotesize  Fig. 3. Left Panel: The cumulative mass functions computed at $z=6$ for SD sterile neutrino models with $m_S = 60$~GeV and $m_{\nu}=7.1$~keV, for different values of the Yukawa coupling $y$ (shown by the labels on the right) in the freeze-in limit. The shaded areas correspond to the observed number density of galaxies with $M_{\rm UV}\leq -12.5$ within within 1-$\sigma$, 2-$\sigma$, and  3-$\sigma$  confidence levels. Right Panel: The maximum value $\overline{\phi}$ (including the theoretical uncertainties) of the predicted number density of  DM halos at $z=6$ for the case with $m_{\nu}=7.1$~keV as a function of the Yukawa coupling $y$. The upper shaded areas represent the observed number density of galaxies with $M_{\rm UV}\leq -12.5$ within 1-$\sigma$, 2-$\sigma$, and  3-$\sigma$ confidence levels. }

For each value of $m_S$, we compute the power spectrum corresponding to each point in the $\lambda-y$ plane as described in Sect.~\ref{sec:DMModels}. Then we represent in Fig. 4, left panel, the regions of the parameter space consistent with the galaxy number densities measured in the HFF ($\overline{\phi}\geq \phi_{\rm obs}$). These  regions clearly split into a freeze-out (for $\lambda\geq 10^{-6}$) and  freeze-in (for $\lambda\ll 10^{-6}$) family.  For the freeze-out family, decreasing the scalar mass  $m_S$ leads to tighter bound on $y$, while yielding an approximate lower bound of $\lambda\gtrsim 10^{-5.2}$ for the Higgs portal coupling. For the freeze-in family, decreasing the scalar mass $m_S$ pushes the admitted values of $\lambda$ to progressively smaller values, while providing progressively stronger limits on $y$. Note that, moving from $m_S=65$~GeV to $m_S=60$~GeV, the allowed freeze-in region shifts to appreciably smaller values of $y$ (actually out of the plot window). This is because a scalar $S$ with 60~GeV has a mass $m_S<m_h/2$. This opens up an entirely new production channel, namely Higgs decay $h \rightarrow SS$, which actually dominates the production rate. In this case the scalars (and hence the sterile neutrinos) are produced with a much larger number density once the Higgs decay sets in. Thus, for the same combination of $\lambda-y$, this larger number density requires a smaller sterile neutrino mass, yielding larger velocities for the DM particles and suppressing the abundance of low-mass halos. 

\begin{center}
 \hspace{-0.4cm}
 \begin{tabular}{lr}
 \includegraphics[width=0.51 \textwidth]{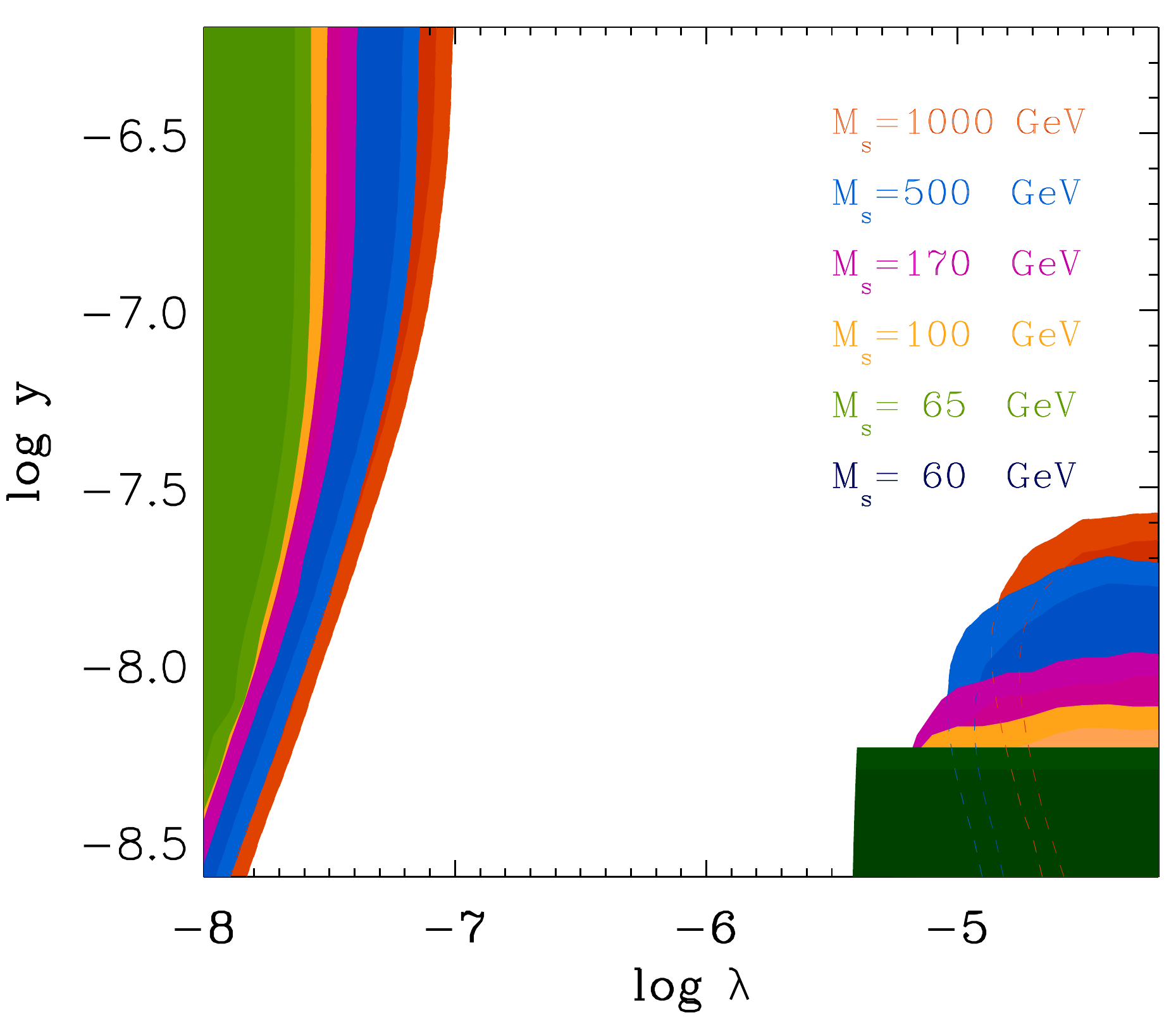} & \includegraphics[width=0.45 \textwidth]{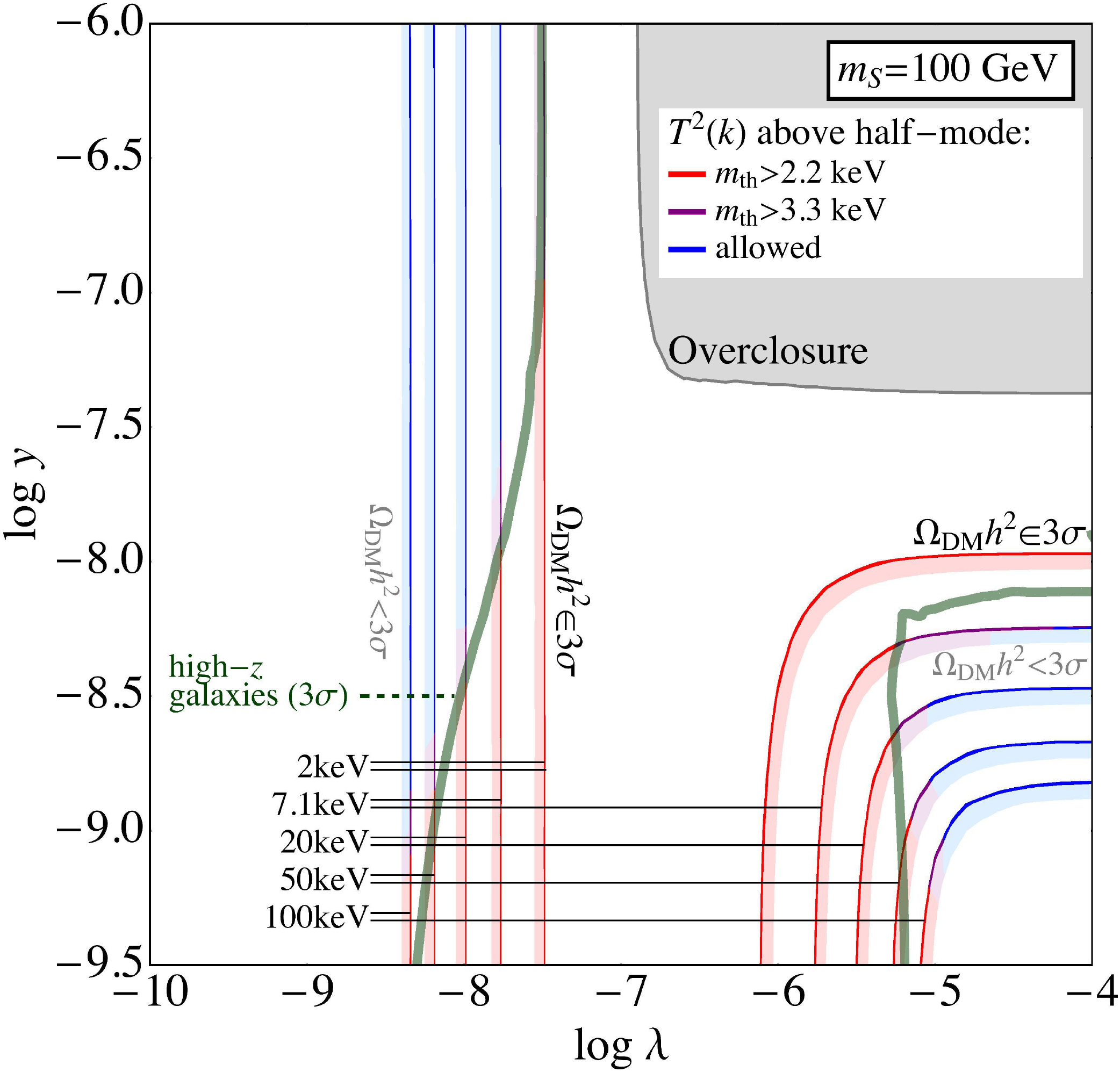}
 \end{tabular}
 \end{center}
 \vspace{-0.5cm}
{\footnotesize  Fig. 4. Left Panel: the constraints on the parameter space of SD production model for sterile neutrino. The colored regions represent the allowed region ($\overline{\phi}\geq {\phi}_{obs}$) in the $\log \lambda$--$\log y$ plane for each value of the  scalar mass $m_S$ (shown by different colors according to the labels on the right). For each scalar mass (and hence for each color) lighter and darker tonalities represent the 1-$\sigma$ and the 2-$\sigma$ confidence levels, respectively. Right Panel: For the case $m_S=100$ GeV, we compare our bounds from HFF galaxies derived in the present work (green thick lines) with a version of Lyman-$\alpha$ bounds derived by \cite{Konig:2016dzg}: the latter are represented as red (forbidden cases), blue (allowed cases), and pruple (constrained cases) segments in the lines of constant $m_{\nu}$ values in  the log $\lambda$-log $y$ plane.  }

Finally, the right panel of Fig. 4 displays an explicit comparison of our high-$z$ HFF bounds for the case $m_S=100$ GeV with a version of Lyman-$\alpha$ bounds derived by \cite{Konig:2016dzg}.  The latter authors of have introduced an approximate method of generalizing Lyman-$\alpha$ bounds to non-thermal cases, by computing the ratio of power spectra of the case under consideration to CDM and demanding the ``upper'' (i.e., high power) part of the spectrum to be allowed when compared to a certain bound. Using the thermal relic mass bounds of $2.2$ and $3.3$~keV, the blue (consistent with both bounds), purple (only consistent with the first bound), and red (inconsistent with both bounds) regions displayed in the right panel of Fig. 4 can be derived. Our HFF-bound, in turn, is displayed by the green line. As one can clearly see, the constraints from HFF galaxies  basically track the Lyman-$\alpha$ bounds: although the latter are slightly stronger, they may possibly suffer from unknown systematics, given that the distribution of the intergalactic material (IGM) and not that of the DM is observed. However, the fact that both types of bounds are basically tracking each other adds significant credibility to both constraints.

\subsection{\label{sec:Results:Fuzzy}Fuzzy DM}

The large observed number density  of high redshift galaxies turns out to provide particularly strong constraints on Fuzzy DM, i.e., scenarios based on wavelike DM composed by ultra-light bosons. In this case, we compute the cumulative halo mass function directly from the formula in \equref{eq:HaloMassDerivativeFuzzy} derived by~\cite{Schive:2015kza} from dedicated $N$-body simulations. This depends on the single free parameter constituted by the particle mass $m_\psi$, so that the exploration of the parameter space is straightforward. In the left panel of Fig. 5 we show the cumulative halo mass function for different values of the DM particle mass (in units of $10^{-22}$~eV), where we have conservatively increased  by 10\% the model number densities derived from \equref{eq:HaloMassDerivativeFuzzy} to account for theoretical  uncertainties~\citep[see][]{Schive:2015kza}. The strong suppression in the number of low-mass halos compared to the CDM case yields a lower limit $m_\psi\geq 10^{-21}$ eV for the DM particle mass (right panel of Fig.5) at  3-$\sigma$ confidence level. This tightens by a large factor the previous bounds on $m_\psi$ derived from the luminosity function of high-redshift galaxies in the Hubble Deep Field ($m_\psi\geq 1.2\times10^{-22}$~eV) given by~Schive et al. (2016; see also Corasaniti et al. 2016). This is due to the increase (by more than one order of magnitude) in the number density of galaxies achieved by  HFF observations of faint galaxies due to the magnification of the foreground clusters, which allowed to push the detection limit at $z=6$ from $M_{\rm UV}=-15$ to $M_{\rm UV}=-12.5$. 

Our results constitute the \emph{tightest constraint on Fuzzy DM particles derived so far}, and have a strong impact for the whole class of models based on Fuzzy DM. In fact, even allowing for observational and numerical uncertainties, all results in the literature indicate that the mass of Fuzzy DM particles should be in the range $m_\psi=(1-5.6)\cdot 10^{-22}$~eV to provide solitonic cores matching the observed  density profile of nearby dwarf galaxies~\citep{Schive:2014dra,Marsh:2015wka,calabrese2016,lora2012,gonzalez2016}. This is inconsistent at more than 3-$\sigma$ confidence level with our lower limits, strongly disfavoring such a scenario as a viable solution to the cusp-core problem of dwarf galaxies.

\begin{center}
  \includegraphics[width=0.49\textwidth]{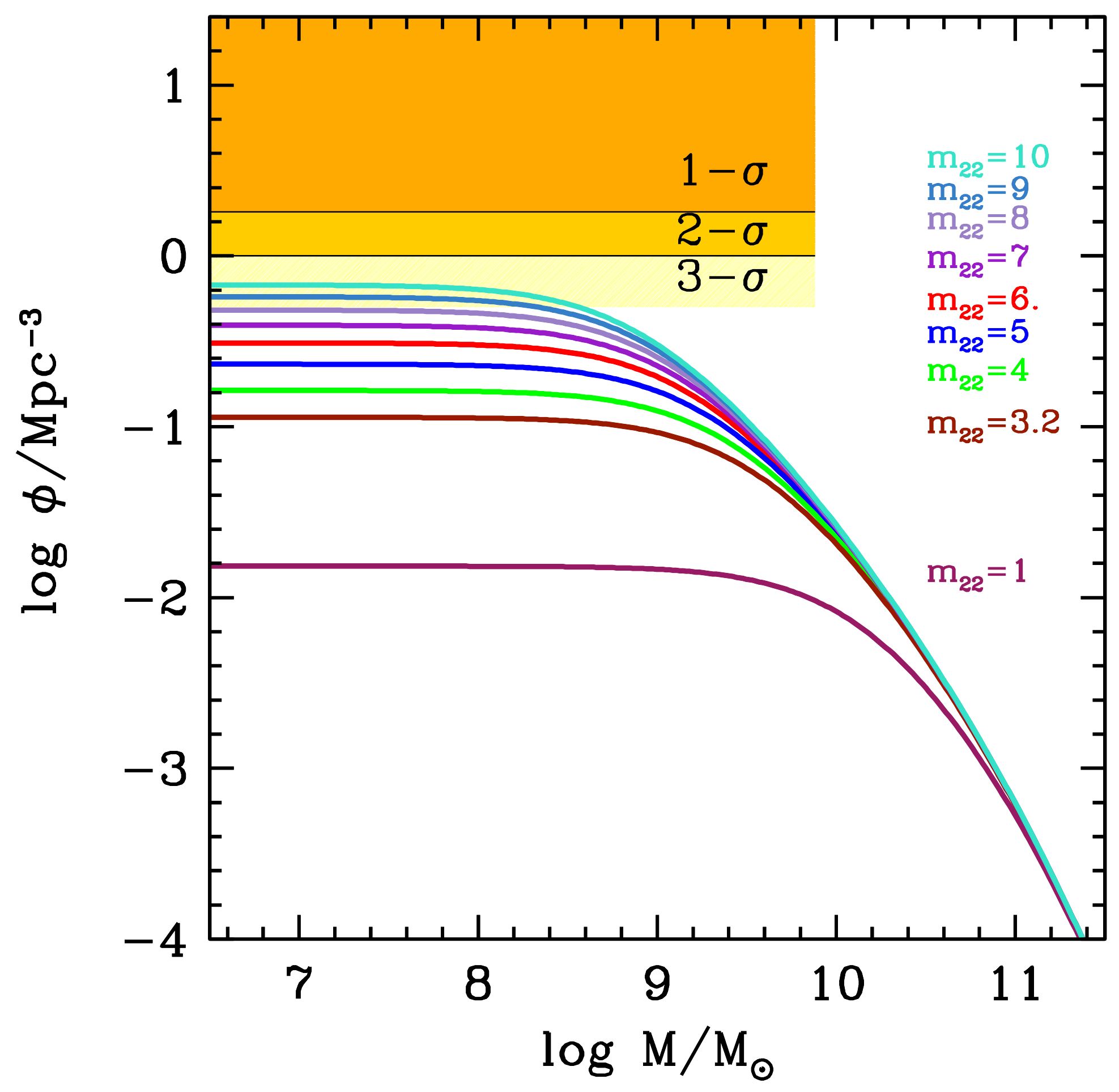}
  \includegraphics[width=0.50\textwidth]{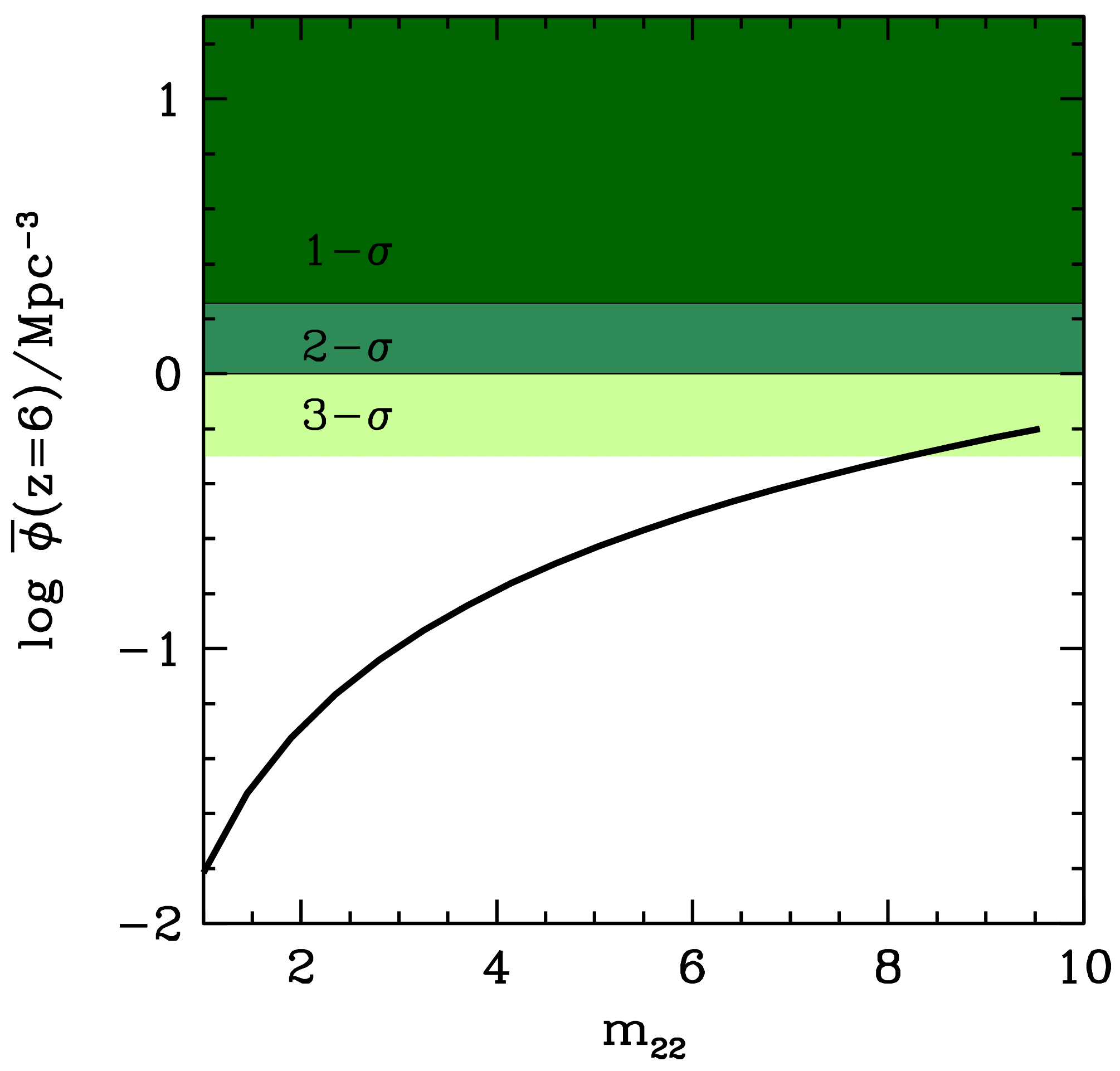}
\end{center}
\vspace{-0.5cm}
{\footnotesize  Fig. 5. Left Panel: The cumulative mass functions computed at $z=6$ for Fuzzy DM models with different values of the wavelike DM particle mass $m_{22}=m_\psi/10^{-22}\,{\rm~eV}$. The shaded areas correspond to the observed number density of galaxies with $M_{\rm UV}\leq -12.5$ within within 1-$\sigma$, 2-$\sigma$, and  3-$\sigma$  confidence levels. Right Panel: The maximum value $\overline{\phi}$ (including the theoretical uncertainties) of the predicted number density of  DM halos $\phi$ at $z=6$ as a function of $m_{22}$. The upper shaded areas represent the observed number density of galaxies with $M_{\rm UV}\leq -12.5$ within 1-$\sigma$, 2-$\sigma$, and  3-$\sigma$  confidence levels.}

\section{\label{sec:Conclusions}Conclusions and Discussion}

We have shown that the Hubble Frontier Field measurements of the abundance of ultra-faint lensed galaxies at $z\approx 6$ down to faint magnitudes $M_{UV}\approx -12$ have profound implications on  the nature of DM, in particular for DM models based on~keV sterile neutrinos and on ``fuzzy'' Dark Matter models
 considered in the present work.

For the case of resonant production of sterile neutrinos, the high galaxy number densities $\sim 1/$ Mpc$^3$ measured at $z=6$ by~\citep{Livermore:2016mbs} provide stringent limits on the combination of sterile neutrino mass $m_{\nu}$ and mixing parameter $\sin^2(2\theta)$. In particular, our method provides lower limits on the values of $\sin^2(2\theta)$ for different values of $m_{\nu}$ which are {\it complementary} to the upper limits derived by the non-observations of decay lines in the X-ray spectra of dwarf galaxies~\citep{Sekiya:2015jsa,Jeltema:2015mee,Riemer-Sorensen:2014yda,Adhikari:2016bei},  thus restricting the allowed portion of the parameter space of such models to a narrow region; e.g., the region around the potential 3.5~keV line ($m_{\nu}=7.1$~keV) is still allowed, but restricted to a narrow range of values for $-11.4\leq\log \sin^2(2\theta)\leq-10.2$. The constraints we obtain on $\sin^2(2\theta)$ for different values of $m_{\nu}$ are close to those obtained from the abundance of low-mass satellites of the Milky Way (Schneider 2016) for a wide range of values $m_{\nu}\lesssim 10$~keV. This  supports the reliability of methods based on structure formation as probes for DM models with suppressed power spectra, especially considering that the two methods are applied to galaxies observed at cosmic times differing by $\sim 13$~Gyr. However, the approach in the present paper is not  affected by the uncertainties affecting methods based on the abundance Milky Way satellites, like the assumed mass of galaxy halos and sub-halos, or the assumed isotropic distribution of satellites. On the other hand, for large sterile neutrino masses of $m_{\nu}\approx 10$~keV, the abundance of high redshift galaxies seems to provide  tighter limits on $\sin^2(2\theta)$ compared to the abundance of Milky Way satellites. This is related to the larger sensitivity of the former method to the shape of the power spectrum at large wavenumber $k$. In turn, this is due to the different dependence on the variance $\sigma^2 (M)$ in the mass function in \equref{eq:HaloMassDerivative} compared to the mass distribution of satellite sub-halos, and to the minimal mass assumed for satellite halos in the Milky Way  ($M\geq 10^8 M_{\odot}$), which excludes tiny halos as hosts of satellite galaxies, thus reducing the sensitivity of such a method with respect to very large $k$-modes. Regarding the limits from Lyman-$\alpha$ absorption lines of distant quasar spectra, the present constraints are  less stringent than those obtained by \citet{Schneider:2016uqi}. It is however important to notice that our method is entirely independent of the baryon physics entering the Lyman-$\alpha$ method, which could affect the thermal state of the intergalactic medium~\citep[see][]{Viel:2013apy,Garzilli:2015iwa}.

For the case of sterile neutrino production via scalar decay, our method provides constraints on the combination of scalar mass $m_S$, Higgs portal coupling $\lambda$, and Yukawa coupling $y$, which are very close to those derived in~\cite{Konig:2016dzg} -- as shown in Fig. 4 for the case $m_S=100$~GeV.  In the  freeze-in limit (see Section 3)  of small Higgs portal coupling $\lambda\ll 10^{-6}$, where the Yukawa coupling constitutes the leading quantity in the determining the model properties, this is due to the strong dependence of the predicted maximum halo number density $\overline{\phi}$ on $y$ (shown in the right-hand panel of Fig. 3 for the case of sterile neutrino mass $m_{\nu}=7$~keV). When compared to the observed high-redshift galaxy abundance this allows to set a constraint $y \geq 9\times 10^{−9}$  at 2-$\sigma$ confidence level.

As for the Fuzzy DM, the results from our method have a relevant impact for the whole class of such models. In fact, extending the analysis by~\cite{Schive:2015kza} to compare with the recent measurements of galaxy abundance from the Hubble Frontier Fields yields unprecedented limits on the mass of candidate DM particles of $m_{\psi}\geq 1.2 \times 10^{-21}$ eV at 2-$\sigma$ confidence level and of $m_{\psi}\geq 8 \times 10^{-22}$ eV at 3-$\sigma$. These values are one order of magnitude larger than those required to yield core sizes of dwarf galaxies large enough to match the observations~\citep{Marsh:2015wka}. Indeed, such limits for the Fuzzy DM mass would correspond to an upper limit $M_J \lesssim 3\cdot 10^5\,M_{\odot}$ (see Sect. 3.3) for the mass scale associated with the de Broglie wavelength of wavelike DM, pushing the scale where the "fuzzy" nature of such DM shows up to values much smaller than those involved in galaxy formation. This result is consistent with the independent constraints derived from the phase space density of local galaxies  by~\cite{devega2014}. 

The bounds presented here are based solely on the statistical properties of DM halos. Indeed,  tighter limits can be obtained adopting a physical description of the baryonic processes involved in galaxy formation, as shown in \citet{menci2016a}. This is the approach taken, e.g., by Corasaniti et al. (2016) who compared the observed luminosity function to theoretical luminosity functions derived from 
the abundance matching technique applied to high-resolution N-body simulations.  The uncertainties associated to baryon physics introduced in such approach can be suppressed by complementing the statistical analysis with the study of relations among galaxy properties that are specific to each DM scenario and as such carry additional information on the nature of DM that is potentially testable with observations.

While our method is robust and independent of the baryon physics entering galaxy formation, and does not rely on any estimate for the DM mass of the observed galaxies, the above results are based on the abundance of $z=6$ galaxies measured in the HFF by \cite{Livermore:2016mbs}. The number density derived by \cite{Livermore:2016mbs} is based on 167 galaxies at $z\ge 6$, and it is thus robust from the statistical point of view; in any case, the statistical uncertainties have been considered in computing the confidence levels of our results. There are however subtle systematic effects related to the estimation of the survey volume, i.e., the variance of the lensing magnification maps of HFF clusters and the physical sizes of faint, high-z galaxies which are strongly magnified by the cluster potential well. 

These effects have been critically analyzed by \cite{bouwens16_magn}, who showed that they could result into luminosity functions flatter than that derived by \cite{Livermore:2016mbs} at $M_{\rm UV}\ge -14$.
 As discussed in Sect. 2.1, the different estimates of the magnification uncertainties obtained by Bouwens et al. (2016) would yield looser limits on the parameter space of DM models. E.g., an observational number density $log\,{\phi}_{obs}/{\rm Mpc}^{-3}=-0.62$ corresponding to the luminosity function by Bouwens et al. (at 2-$\sigma$ confidence level, see Sect. 2.1) would lead to $log \sin^2\left(2\theta\right)\geq -10.2$ for $m_{\nu}$=5 keV, to $log \sin^2\left(2\theta\right)\geq -11.5$  for $m_{\nu}$=7 keV, and to $log \sin^2\left(2\theta\right)\geq -12.2$  for $m_{\nu}$=10 keV  in the case of resonantly produced sterile neutrinos, thus lowering the lower bounds in fig. 2. 
For the Scalar Decay Models of sterile neutrinos, the lower limit on $y$ in fig. 4 would be decreased to $y\geq 5\times 10^{-8}$ (for the case 
$m_{\nu}$=7 keV), while for the case of Fuzzy DM, adopting the Bouwens et al. (2016) luminosity functions would yield $m_{\psi}\geq  5\times 10^{-22}$ eV at 2-$\sigma$ confidence level. 

 Thus, on the observational side, the first step to improve the results presented in this paper consists in a deeper understanding of the systematics associated with the lensing observations of faint, high-redshift galaxies.
As for the lensing magnification, with typical values $\mu>10$ and as large as $\sim 50-100$ for the faintest galaxies, the analysis in  \cite{Livermore:2016mbs} shows that large differences are found in the magnification estimates of individual galaxies when different lensing models are assumed. This is in agreement with the recent analysis by \cite{meneghetti2016}, who compared the performances of several lensing models on artificial images mimicking the depth and resolution of HFF data. They find that the largest uncertainties in the lens models are found near sub-structures and around the cluster critical lines, concluding that uncertainties in the magnification estimate are growing with the magnification value itself (30\% of magnification uncertainties at $\mu\ge 10$). However, while the magnification of each individual galaxy could be subject to a large variance, the aggregated information of the abundance of high-z galaxies is less subject to such systematics. The analysis in \cite{Livermore:2016mbs} shows that the different lensing models introduce minimal changes in the best-fit luminosity functions, even at the faintest absolute magnitudes $M_{\rm UV}=-12.5$  analyzed at $z\sim 6$, with a systematic uncertainty on the slope $\alpha$ of the luminosity functions at $z\sim 6-7$  below 2\%. Although such an uncertainty would not change appreciably the results presented in this work, the proper procedure to be adopted in deriving the variance in the luminosity function due to the magnification is still matter of debate \citep[see ][]{bouwens16_magn}. 

The other systematic effect affecting the  luminosity functions measured by \cite{Livermore:2016mbs} is related to the size distribution of high-$z$ galaxies. In fact, the detection of faint galaxies is strongly affected by their surface brightness distributions, with compact galaxies more easily detected compared to the extended, low -surface brightness ones \citep{grazian2011}. In the simulations carried out by \cite{Livermore:2016mbs}, a normal distribution of half-light radii $r_h$ has been assumed with a peak at 500 pc. While similar (or even higher) number densities are obtained for values $r_h\gtrsim 100$~pc, for smaller values a significant suppression of the faint-end logarithmic slope $\alpha$ of luminosity function is found \citep[up to 10\% in the case $r_h= 40$ pc, ][]{bouwens16_size}. While \cite{bouwens16_size} find that the typical values of $r_h$ at $z=6$ is of the order of $\sim 25-80$~pc for $M_{\rm UV}\ge -15$, larger values are usually obtained in the literature: extrapolating the size distribution obtained by \cite{kawamata2015} for $M_{\rm UV}\sim -20$ galaxies at $z\sim 6$ to fainter magnitudes  \citep[assuming a size-luminosity relation of $r_h\propto L^{0.5}$][]{grazian2012}, the typical sizes of galaxies with $M_{\rm UV}=-12.5$ range from $\sim$ 40~pc  to $\sim 100$~pc with a log-normal distribution; \cite{laporte2016} analyzed the size distribution of $z\sim 7$ Lyman-break galaxies finding a typical value of 250~pc for galaxies with  $-19\leq M_{\rm UV}\leq -17$. Thus, while the completeness correction adopted by \cite{Livermore:2016mbs} is consistent with most of the extrapolations in the literature, assessing the actual size distribution of ultra-faint galaxies at $z\approx 6$ would constitute an important improvement for a precise determination of the number density of such galaxies, so as to  lower this source of systematics below the present value $\approx 10\%$. In the future we plan to analyze the impact of the size-luminosity relation on the HFF luminosity function based on detailed simulations, using the same technique adopted in \cite{grazian2011,grazian2012} to comply with the delicate issues (i.e., distortions due to the strong lensing amplification) involved in the measurement of sizes for such faint, noisy galaxies.

Thus, although the measurement by \cite{Livermore:2016mbs}  constitutes a state-of-the-art achievement, the analysis of the HFF observations is open to several advancements. This is true not only for the most magnified objects (as discussed above), but also for those moderate magnifications ($\mu \lesssim 10$), for which the largest source of uncertainty is constituted by the accuracy of photometric redshifts. In this respect, improvements can be readily obtained through the combination of different procedures \citep[e.g.,][]{2016A&A...590A..31C}, an approach which is known to reduce systematic effects of photo-$z$ estimates \citep{dahlen2013}. Indeed, deep MUSE observations \citep{vanzella2016}  are already able to provide the spectroscopic redshifts for faint $z\ge 6$ galaxies in the HFF pointings, improving both the lensing magnification maps and the photometric redshifts. Finally, the analysis in \cite{Livermore:2016mbs} is based at present only on the first two fields of the HFF survey: the inclusion of the remaining four strong lensing clusters \citep{lotz2016} will reduce both statistical uncertainties and mitigate possible cosmic variance effects. In addition, it is not to be disregarded that the six available parallel HFF pointings at depth comparable to the HUDF will improve the determination of the logarithmic faint-end slope of the luminosity function that can work as a valuable baseline for interpreting the abundance of fainter lensed sources. In a few years from now a significant leap will be made possible by the availability of deep JWST imaging. In particular, the capability of reaching 30.5~AB (at $S/N=5$) in deep NIRCam fields \citep[e.g.,][]{2015arXiv151204530F} will improve by~1.5 mags the depth of current HFF imaging, reaching absolute magnitudes of $M_{\rm UV}\approx -11$, and will yield 5 times larger samples of high-redshift galaxies \citep{Laporte:2014mca} while significantly improving photometric selections through the availability of rest-frame optical photometry of high-$z$ sources.

\begin{acknowledgements}
We  thank the referee for constructive comments that helped to improve the paper.\\ 
\end{acknowledgements}



\end{document}